\documentstyle[preprint,aps]{revtex}
\begin{document}
\draft

\title{Ferroelectric and Dipolar Glass Phases of Non-Crystalline Systems}

\author{
G. \ Ayton$^\dagger$, M.\ J.\ P.\ Gingras$^{*,**}$, and G.\ N.\ Patey$^\dagger$}

\address{
$^\dagger$Department of Chemistry, University of British Columbia, Vancouver,
British Columbia, Canada V6T 1Z1 }

\address{
$^*$TRIUMF, 4004 Wesbrook Mall, Vancouver, British Columbia, Canada V6T
2A3}

\date{\today}

\maketitle

\begin{abstract}
\setlength{\baselineskip}{0.6cm}

In a recent letter [Phys. Rev. Lett. {\bf 75}, 2360 (1996)] we briefly
discussed the existence and nature of ferroelectric order in positionally
disordered dipolar materials. Here we report further results and give a
complete description of our work. Simulations of randomly frozen and
dynamically disordered dipolar soft spheres are used to study ferroelectric
ordering in non-crystalline systems. We also give a physical interpretation of
the simulation results in terms of short- and long-range interactions. Cases
where the dipole moment has 1, 2, and 3 components (Ising, XY and XYZ models,
respectively) are considered. It is found that the Ising model displays
ferroelectric phases in frozen amorphous systems, while the XY and XYZ models
form dipolar glass phases at low temperatures. In the dynamically disordered
model the equations of motion are decoupled such that particle translation is
completely independent of the dipolar forces. These systems spontaneously
develop long-range ferroelectric order at nonzero temperature despite the
absence of any fined-tuned short-range spatial correlations favoring dipolar
order. Furthermore, since this is a nonequilibrium model we find that the
paraelectric to ferroelectric transition depends on the particle mass. For the
XY and XYZ models, the critical temperatures extrapolate to zero as the mass
of the particle becomes infinite, whereas, for the Ising model the critical
temperature is almost independent of mass and coincides with the ferroelectric
transition found for the randomly frozen system at the same density. Thus in
the infinite mass limit the results of the frozen amorphous systems are
recovered.

\end{abstract}


\setlength{\baselineskip}{0.8cm}

\narrowtext
\section{Introduction}

Computer simulations of fluids of strongly interacting dipolar spheres have
established the existence of a stable ferroelectric liquid crystal phase
\cite{wei,weis}. The phase is truly fluid in that it exhibits long-range
orientational (ferroelectric) order, but only short-range spatial
correlations. Interestingly, the local spatial correlations found in these
ferroelectric fluids were similar to those in the ferroelectric tetragonal-I
lattice, which is the lattice structure believed to be the low-temperature
ferroelectric solid phase for dense dipolar hard spheres \cite{tao}. The
translational mobility of the particles in the fluid phase allowed specific
short-ranged correlations to build up, and the system spontaneously polarized.
>From these results, it was believed that the establishment of a ferroelectric
liquid phase was largely driven by well-tuned specific short-range spatial
correlations \cite{gingras-holdsworth}.

On the other hand, experimental systems of near spherical Fe$_3$O$_4$ magnetic
particles in a frozen non-magnetic solvent \cite{hochli,jonsson,vug}, which
interact via magnetic dipole moments, do not exhibit a ferroelectric phase
upon cooling, but rather, develop magnetic irreversibilities similar to the
situation in random magnetic systems known as spin glasses \cite{BYFH}. In
these systems, the particles are frozen at random locations at all
temperatures, but above a certain temperature the particle dipoles can freely
rotate \cite{free_rotation} The simplest interpretation of these results is
that the lack of fined-tuned spatial correlations inhibit the formation of
long-range ferroelectric order, while the large degree of randomness in the
positions lead to random frustration for the dipolar interactions and,
consequently, to the formation of a ``dipolar glass'' at low temperature. The
anisotropic nature of the dipolar interaction would lead one to expect that
the glass transition in these frozen ferrofluids is of thermodynamic (as
opposed to purely dynamic) origin, and hence occurs at a nonzero glass
transition temperature, $T_g$, characterized by a divergent spin-glass
susceptibility \cite{BYFH,gingras_sg}.

To summarize, the naive picture that emerges is that some crystalline lattices
(e.g., body-centered cubic, face-centered cubic, tetragonal-I) and some
dipolar fluids can spontaneously develop long-range ferroelectric order
\cite{restric_to_FE} because they have ``suitable'' spatial correlations. On
the other hand, low-density randomly frozen dipolar systems lack these
correlations, exhibit random frustration, and therefore have dipolar glass
ground states \cite{vdw-example}. Also, we would expect that dilution of the
dipoles on a crystalline lattice, which displays ferroelectric order for
full occupation, will eventually lead to a transition from ferroeletric order
to dipolar glass below a critical occupation density, again due to the
increase build up of random frustration with decreasing dipole density
\cite{hochli,xu_thesis}.

However, in recent papers, Zhang and Widom \cite{widom,zw95} proposed a mean
field theory that predicts ferroelectric phases in dipolar systems that
lacked any specific spatial correlations, provided the density of the
particles, $\rho$, was above a critical value, $\rho_c$. They considered
amorphous solids of dipolar hard spheres where the particles were free to
rotate, but were frozen at random sites. Specifically, they assumed complete
randomness where the radial distribution function, $g(r)$, describing the
probability that two particles are separated by a distance $r$, was set to
$g(r)=1$ for $r > \sigma$, where $\sigma$ is the diameter of the sphere. Their
prediction of ferroelectric phases in dipolar systems that lack any specific
spatial correlations suggests that well-tuned short-range structure may not be
necessary for ferroelectric phase formation. Although the experimental results
on frozen ferrofluids seem to contradict this assertion, Zhang and Widom
suggested that the experimental dipole density was possibly too low (i.e.,
below $\rho_c$) for ferroelectric order, and hence a dipolar glass state was
found.

The role of short-ranged spatial correlations on ferroelectric phase formation
is still not well understood. The question of whether or not dense spatially
disordered dipolar materials can have ferroelectric phases was briefly
discussed in Ref. \cite{us}. In this paper, the effect of spatial disorder on
the phase behavior of dense dipolar systems is investigated with molecular
dynamics (MD) and Monte Carlo (MC) simulations. Systems where the dipole
vector has one, two and three components are considered and we refer to these
as the Ising, XY and XYZ models, respectively. The fundamental forces that
promote and destroy ferroelectric order in dipolar systems are identified
and discussed.

The remainder of this paper is organized as follows. In Section II we briefly
discuss the generic temperature vs density phase diagram one might expect to
find for diluted ferroelectric lattices and amorphous dipolar systems. In
Section III, the models and simulation methods employed are described. Section
IV is concerned with ferroelectric and dipolar glass ordering found in
amorphous frozen dipolar systems. A new simulation technique devised to study
orientational order in spatially random media is introduced in Section V. The
results  of these ``dynamically disordered'' simulations offer significant
insight into the observed behavior of both dipolar fluids \cite{wei,weis} and
the dipolar amorphous solids \cite{hochli,jonsson,vug,widom}. A mean field
theory which describes the ferroelectric transition when only reaction field
interactions are present is given in Section VI. This helps us understand the
competition between the long-range reaction field interactions and
shorter-ranged contributions to the energy. This competition is a key feature
determining whether or not a system displays ferroelectric order. Finally, our
main results and conclusions are summarized in Section VII.

\section{Temperature-Density Phase Diagram in Random Dipolar Systems}

It is useful to briefly discuss the temperature, $T$, vs density, $\rho$,
phase diagram (sketched in Fig. 1) that one might expect to observe in random
dipolar systems based on our current understanding of randomly frustrated
magnetic spin glasses \cite{hochli,BYFH,xu_thesis,x-rho_pd}.

First, consider a Bravais lattice (e.g., body-centered cubic) where all sites
are occupied by an $n$-component classical dipole such that $\rho=\rho_{\rm
max}$. We will assume here that the positions of the particles at the lattice
sites are fixed. At high temperature, the system is in a disordered
paraelectric phase (see Fig. 1). For a perfect lattice (all sites occupied), a
transition to a long-range ferroelectric phase occurs at some critical
temperature, $T_c$ \cite{xu_paper}. As the system is diluted, a ferroelectric
phase remains observable for sufficiently large $\rho>\rho_c$, but with $T_c$
decreasing as more sites are vacated and $\rho$ decreases. As well, the
zero-temperature polarization, $P(T=0)$, decreases from its full value at
$\rho=\rho_{\rm max}$ as $\rho$ approaches $\rho_c$ from above. Based on work
on spin glasses \cite{BYFH,x-rho_pd}, we expect that upon cooling within the
ferroelectric phase, there will be a low-temperature ``mixed'' state ($\rho_c
< \rho < \rho_{\rm max}$) where strong irreversibilities and glassy behavior
develops (i.e., the polarization in field-cooled and zero field-cooled
experiments differs below the long-dashed line in Fig. 1, which is commonly
referred to as the Almeida-Thouless line in the mean-field theory of Ising
spin glasses \cite{BYFH}). However, long-range ferroelectric dipolar order is
not lost in the mixed phase and there is likely no decrease of the
polarization upon cooling from the ferroelectric state down into the mixed
state \cite{x-rho_pd}. For $\rho<\rho_c$, the random frustration leads to a
transition to an unpolarized dipolar glass state characterized by an
Edwards-Anderson order parameter \cite{BYFH}. Most naively, we expect that the
intrinsic anisotropy of dipolar interactions stabilizes a dipolar glass phase
at nonzero temperature in the three-dimensional Ising spin glass universality
class \cite{gingras_sg} (with the proviso that recent numerical work suggests
that the existence of a thermodynamic spin-glass transition in the
three-dimensional Ising spin glass is {\it still} not completely settled
\cite{young_unsettled}.). In diluted magnetic systems with short-range
frustrated interactions, such as diluted CdMnTe and  EuSrS, one finds that the
glass transition temperature vanishes for a nonzero density, $\rho_-$, which
corresponds to a site percolation threshold \cite{BYFH}. For classical
dipoles, $\rho_-$ is zero due to the long-range nature of the interactions.
However, for small non-classical dipole moments, quantum effects will likely
move $\rho_-$ to a finite value due to the random-transverse fields acting on
each spin in the frozen state \cite{blinc}.

Consider now the case of amorphous dipolar systems where there are little or
no spatial correlations among the locations of the dipoles (i.e., the dipoles
are not on lattice sites)~\cite{jonsson}. Recently, Zhang and Widom
\cite{widom,zw95} have argued using a mean field model that frozen amorphous
dipolar systems could display spontaneous ferroelectric order for a packing
density $\rho >\rho_c(n)$ for $n=1$ (Ising model) and 3 (XYZ model) component
dipoles \cite{widom,zw95}. It should be noted here that one clearly cannot make
arbitrarily dense amorphous systems lacking {\it all} spatial correlations.
However, the critical density for ferroelectric order in amorphous Ising and
XYZ systems was found in Ref.~\cite{widom,zw95} to be reasonably less than the
close-packed value, and it should be possible to test the theoretical
predictions for a range of densities. In other words, varying the temperature
for sufficiently large $\rho$, one should according to Zhang and Widom follow
the trajectory $a)$ in Fig. 1, and find a ferroelectric phase below
$T_c(\rho,n)$.

The main motivation of the present study was to investigate this possibility.
Below, we report results from extensive computer simulations which show that
for a high density amorphous system, the Ising model does display a
ferroelectric transition. In fact, for the density considered, the Ising
system developed a polarization close to the full maximum value (i.e., the
Ising system considered was at $\rho\gg \rho_c)$. However for the same and
even larger densities, we found that the XY and XYZ models showed instead a
dipolar glass transition and no ferroelectric order was observed. Thus, these
systems followed path $b)$ in the phase diagram sketched in Fig. 1.
 
\section{The Model and Simulation Details}

We consider systems of point dipoles embedded at the center of soft spheres.
Soft spheres (ss) are defined by the pair potential
\begin{equation}
u_{{\rm ss}}(12) = 4\varepsilon(\sigma/r)^{12}\ ,
\end{equation}
where the parameters $\varepsilon$ and $\sigma$ are the fundamental units of
energy and length, and $r$ is the distance between the particles. The
dipole-dipole (DD) interaction is given by
\begin{equation}
u_{DD}(12) = -3(\bbox{\mu}_{1} \cdot {\bf r} )(\bbox{\mu}_{2} \cdot
{\bf r})/r^{5}
+ \bbox{\mu}_{1} \cdot \bbox{\mu}_{2}/r^{3}\ ,
\end{equation}
where $\bbox{\mu}_{i}$ is the dipole of particle $i$ and ${\bf r}$ is the
interparticle vector. As noted above, we consider Ising, XY and XYZ models.
In all three cases the potential is given by Eq. (2).

All calculations were carried out employing periodic boundary conditions and
the long-range dipolar forces were taken into account using Ewald summation
methods. The dielectric constant of the surrounding continuum, $\epsilon'$,
necessary in the Ewald method \cite{wei,deleeuw,kusalik} was taken to be
infinity (i.e., conducting boundary conditions). For the present dense
strongly dipolar systems the dielectric constants are sufficiently large that
conducting boundary conditions are an appropriate choice.

The existence of a ferroelectric phase can be detected by calculating the
average polarization per particle, $P$, defined as
\begin{equation}
P =    \frac{1}{N} \langle \sum_{i=1}^{N}
\hat{\bbox{\mu}}_{i}\cdot\hat{{\bf d}} \rangle\ ,
\end{equation}
where $\hat{{\bf d}}$ is a unit vector in the direction of the total
instantaneous moment, ${\bf M}= \sum_{i=1}^{N} \bbox{\mu}_{i}$, and $N$ is the
number of particles in the system. In ferroelectric systems, $P$ is nonzero
and tends to one as the ferroelectric order increases. In a disordered phase,
(either paraelectric or dipolar glass) $P$ will be zero if the system is
sufficiently large. However, in simulations $P$ must be expected to exhibit
significant system size dependence and, as demonstrated below, this must be
carefully checked.

In the present work, we are also interested in the possibility that a system
could orientationally ``freeze'' into an unpolarized state at low temperatures.
Such an orientationally frozen, or dipolar glass phase can be detected by
calculating the root-mean-square (rms) dipole length, $\langle S_{\rm rms} \rangle$, given
by \cite{thompson}
\begin{equation}
\langle S_{\rm rms} \rangle = \frac{1}{N} [\sum_{i}
(<{\bf m}_{i}> \cdot <{\bf m}_{i}>)]^{1/2}\ ,
\end{equation} 
where
$< {\bf m}_{i}> = \tau^{-1}
\sum_{\tau '=0}^{\tau} \hat{\bbox{\mu}}_{i}(\tau ')$,
and $\tau$ is the number of MD timesteps, or MC sweeps (i.e., N attempted
moves). Briefly, for ferroelectric systems both $P$ and $\langle S_{\rm rms} \rangle$ will
be nonzero. If $P \approx 0$ but $\langle S_{\rm rms} \rangle$ is nonzero, the system is an
orientationally frozen dipolar glass. If both $P$ and $\langle S_{\rm rms} \rangle$ are
near zero, the particles are freely rotating as in a plastic crystal or normal
isotropic fluid.

Systems of dipolar soft spheres can be characterized by specifying
the reduced density, $\rho^{*} = N \sigma^{3}/V$, the reduced temperature,
$T^{*} = kT/\varepsilon$, where $k$ is the Boltzmann constant, and the reduced
dipole moment, $\mu^{*} = (\mu^{2}/\varepsilon \sigma^{3})^{1/2}$. In the
present work, the reduced dipole moment and reduced density were constant at
$\mu^{*} = 4$ and $\rho^{*} = 0.8$. This density is well within the range
where Zhang and Widom predict a ferroelectric phase. For example, the lowest
density of the ferroelectric phase is predicted to be $\rho^{*}=0.31$ for the
Ising system and $\rho^{*}=0.55$ for the XYZ model \cite{widom,zw95}. In the
present case with $\mu^{*} = 4$ and $\rho^{*} = 0.8$, the theory of Zhang and
Widom predicts that the Ising model will be ferroelectric if $T^{*} \leq 35.2$
and the XYZ model if $T^{*} \leq 4.8$ \cite{crit}.

\section{Randomly Frozen Systems}

In our simulations the ``randomly'' frozen spatial structure is taken to be a
typical fluid configuration of soft spheres at $T^{*}=10.5$ and $\rho^{*}=
0.8$. Such configurations are readily generated by MD simulation of the
soft-sphere fluid. Unfortunately, it is impossible to have a truly random and
uncorrelated (i.e., the radial distribution function, $g(r)=1$ for $r \gtrsim
\sigma$) spatial configuration at $\rho^{*}=0.8$. The radial distribution
function for a soft-sphere fluid at $\rho^{*}=0.8$ and $T^{*}=10.5$ is shown
in Fig. 2. Clearly, the spatial correlations in this system are weak and very
short-ranged. The point dipoles are located at the centers of the soft
spheres, initially with random orientations. Constant temperature MD
simulations \cite{allen} were carried out for the XY and XYZ models. The
reduced simulation timestep, $\delta t^{*} = (\varepsilon/m\sigma^{2})^{1/2}
\delta t = 0.00125$, where $m$ is the particle mass, was used together with
the reduced moment of inertia $I^{*}=I/m\sigma^{2}=0.025$. Of course, these MD
simulations involved only rotational motion since the particles remain
spatially frozen. Monte Carlo simulations were performed for the Ising model.
One Monte Carlo ``sweep'' consisted of $N$ attempted random flips of the
dipolar orientations. Typically, the results reported involved aging runs of
100,000 MD timesteps or MC sweeps followed by production runs of the same
length.

The average polarization as a function of $1/N$ for the Ising, XY and XYZ
models is shown in Fig. 3. The systems considered ranged from 108 to 864
particles. The polarization values that are plotted were obtained at the
lowest temperature where equilibrium could be achieved independent of the
starting configuration. We call this temperature $T^{*}_{{\rm min}}$. The
values of $T^{*}_{{\rm min}}$ for the Ising, XY and XYZ models are 10.0, 4.0
and 3.5, respectively. Below these temperatures, simulations that were started
from perfectly aligned and random states did not converge to the same result,
at least not in simulation runs of practical length. However, the values of
$T^{*}_{{\rm min}}$ attained are within the temperature range where Zhang and
Widom predict a ferroelectric phase. As discussed below, it appears that the
Ising system develops ferroelectric order with $P>0$ in the thermodynamic
limit. In this case, $T^{*}_{{\rm min}}$  may qualitatively correspond to the
long-dashed line in Fig. 1.

>From Fig. 3 we see that the Ising model at $T^{*}_{{\rm min}}=10.0$ is almost
completely polarized and shows little or no system size dependence. A detailed
$P$ vs $T^{*}$ plot for the frozen $N=256$ Ising system is given below (see
Fig. 7). It can be seen that ferroelectric order develops spontaneously in
this system at $T^{*} \approx 25$. This transition temperature is
significantly lower then that (i.e., $T^{*}=35.2$) predicted by the theory of
Zhang and Widom. Returning to Fig. 3, we see that the XY and XYZ models at
$T^{*}_{{\rm min}}$ show significant polarization for the 108 and 256 particle
systems. However, in both cases the polarization decreases monotonously with
increasing system size and  appears to approach zero in the thermodynamic
limit. The observed polarization of the XY model is strongly dependent on
system size, decreasing from $\sim 0.49$ for $N=256$ to $\sim 0.10$ for
$N=864$. The polarization for XYZ model shows a similar, although not as
pronounced, system size dependence.

Further information about the behavior of the XYZ and XY models can be
obtained by examining the temperature dependence of $\langle S_{\rm rms}
\rangle$. Results for the XYZ model ($N=256$) are shown in Fig. 4. These
results were obtained for samples initially begun with random dipolar
orientations. We see that $\langle S_{\rm rms} \rangle$ is zero at high
temperatures as expected, but becomes nonzero and grows with decreasing
$T^{*}$ at lower temperatures. Similar behavior was observed for the XYZ
system with 864 particles, and for the XY model. The growth of $\langle S_{\rm
rms} \rangle$ at low temperatures without a corresponding development of
polarization provides qualitative evidence that the XYZ and XY models
freeze orientationally to form dipolar glasses.

To summarize, we find no evidence of ferroelectric states in the thermodynamic
limit for  the randomly frozen XYZ and XY models. This clearly disagrees with
the theory of Zhang and Widom, which predicts that the XYZ model should have a
stable ferroelectric phase in the temperature range we consider. Additional MD
calculations at $\rho^{*}=1.05$ were carried out for the XYZ model, but again
no ferroelectric behavior was observed. The Ising model does have a
ferroelectric phase, however, spontaneous polarization was observed at $T^{*}
\approx 25$, which is much lower than the transition temperature predicted by
Zhang and Widom.

\section{Dynamically Disordered Systems}

In the simulations described above, a  configuration was selected from a MD
simulation of soft spheres at $\rho^{*}=0.8$ and $T^{*}=10.5$, and was then
used as a ``typical'' randomly frozen system. Point dipoles were  embedded at
the centers of the soft spheres and rotational MD or MC simulations were
performed. Ideally, as in spin glass models \cite{BYFH}, many (i.e., 100-1000)
randomly frozen configurations should be used to obtain accurate values of the
``disorder-averaged'' polarization. However, for the dipolar systems considered
here this would be a very laborious procedure requiring many long simulations.

As an alternative approach to the study of orientational ordering in random
media, we have used a MD simulation technique where the rotational and
translational equations of motion are completely decoupled. That is, we
consider dipoles embedded in a dynamic random ``substrate'' rather than in a
frozen system. The underlying soft-sphere substrate is a simple fluid. It has
no long-range positional correlations and the short-range correlations are not
influenced by the dipolar interactions (i.e., the spatial correlations are
identical to those of a soft-sphere model [Eq. (1)] at a given density and
translational temperature). However, the ``equilibrium'' state of the dipoles
will depend on the underlying motion of the substrate. This model is similar
in spirit (but not equivalent to due to its lack of obvious energy currents)
to those used in recent studies of non-equilibrium phase transitions in
magnetic systems subject to Levy flights \cite{levy}. It is fundamentally
different from dipolar fluid simulations in that the spatial structure is not
affected by dipole-dipole interactions, but it is also not an amorphous  solid
simulation since the particles move.

With this technique, we can gain a good deal of insight into the role of
short-ranged spatial correlations on phase behavior. By controlling the rate
that the substrate moves relative to the rate of dipolar reorientations, we
can effectively ``turn on'' or ``turn off'' the specific details of the random
spatial structure which the dipoles ``see''. The implementation is
straightforward. The force between two particles is simply given by
\begin{equation}
{\bf f}(12) = -\nabla_{\bf r}u_{{\rm ss}}(12)\ ,
\end{equation}
where $u_{{\rm ss}}(12)$ is the soft-sphere potential defined above. The
torques are given by the dipole-dipole interactions. The spatial structure of
the system is then determined only by the soft-sphere part of the pair
potential. The translational and rotational temperatures are decoupled such
that the structure and motion of the substrate is not affected by changes in
the rotational temperature. The rate at which the substrate moves relative to
the dipolar reorientations can be varied by changing the particle mass.
Obviously, since it is a classical fluid, adjusting the particle mass will
have no effect on the equilibrium structure of the soft-sphere substrate. For
large masses, the substrate changes slowly relative to dipole reorientations.
An extrapolation to the infinite mass case would give results of a randomly
frozen system. For light masses, the substrate moves rapidly relative to
dipole reorientations. The substrate motion may be so rapid
that the dipoles cannot react to structural details and a mean-field-like
limit is reached~\cite{levy}. Thus, from a dipole's point of view, increasing
the mass effectively ``turns on'' the specific structural details of its
surrounding dipolar environment. The moving substrate is a means of simulating
dipolar systems in a dynamically random medium that lacks any specific spatial
correlations. Clearly, in these systems positional correlations which favor
ferroelectric order are not present initially and, due to the decoupling, such
correlations cannot develop as the system evolves. Contact with the randomly
frozen systems can be made by examining the behavior at intermediate masses
and then extrapolating to the infinite mass limit.

In the decoupled MD simulations, the underlying substrate is a
soft-sphere fluid at $\rho^{*}=0.8$ and $T^{*}({\rm translational})=10.5$. All
decoupled simulations were carried out using the reduced timestep, $\delta
t^{*} = (\varepsilon/m^{\prime}\sigma^{2})^{1/2} \delta t = 0.00125$, and the
reduced moment of inertia $I^{*}=I/m^{\prime}\sigma^{2} = 0.025$. The
equations of motion of the soft-sphere substrate were written in terms of the
reduced mass, $m^{*}=m/m^{\prime}$, which could be varied to change the rate
of translational motion of the substrate. Note that $I^{*}$ and the rotational
equations of motion which govern the dipoles have no dependence on $m^{*}$.

$P$ versus $T^{*}$ (rotational) for the XYZ model is plotted in Fig. 5.
Systems with particle masses $m^{*}=1$, 5 and 10 are included. Clearly,
spontaneous polarization develops for all systems and the temperature at which
$P$ begins to rise decreases with increasing mass. For  $m^{*}=5$, the
transition occurs at  $T^{*} \approx 0.55$, and for $m^{*}=10$, at $T^{*}
\approx 0.25$. For $m^{*}=1$, we have also calculated the
Binder ratio, $g_{\rm Binder}$, defined as \cite{BYFH}
\begin{equation}
g_{\rm Binder} = (5/2)-(3/2)\langle|{\bf M}|^4\rangle/\langle|{\bf
M}|^2\rangle^2,
\end{equation}
for systems with 64, 108 and 256 particles. A plot of $g_{\rm Binder}$ vs
$T^{*}$ is included as an inset in Fig. 5. A clear crossing, and hence a
thermodynamic transition, at $T^{*}\approx 1.9$ is evident.

XYZ systems with larger masses (up to $m^{*}=20$) were investigated
but all were disordered above $T^{*}=0.1$. Below $T^{*}=0.1$ calculations for
very large masses converged too slowly to be useful. Similar results were
obtained with $N=108$, 256, and 500 particle systems and the polarization
showed no significant dependence on system size. The results shown in  Fig. 5
strongly suggest that for any finite mass the XYZ model will spontaneously
polarize at some rotational temperature, but as the mass becomes very large
the transition temperature will approach  zero. As shown in Fig. 6, the
dynamically decoupled XY model behaves much as the XYZ system. It can be seen
that systems with $m^{*}=1$, $5$ and $10$ spontaneously polarize at $T^{*}
({\rm rotational}) \approx 6$, $2$, and $1.8$. Again, the transition
temperature decreases with increasing mass.

In the Ising model the potential does not vary as a continuous function of
orientation and hence this model is not well suited to MD simulations.
Therefore, a MC scheme was devised that allowed the substrate to move
independently of the Ising dipoles. This involved combining a soft-sphere MD
simulation with a MC Ising dipole simulation. The soft-sphere substrate
evolved as in the XY and XYZ decoupled calculations, however, after each MD
timestep, $N$ attempted dipole flips (one MC sweep) were performed using the
usual Metropolis Monte Carlo method \cite{allen}. In Fig. 7, $P$ vs $T^{*}$
(rotational) results are plotted for $m^{*}=1$, $m^{*}=5$ and the randomly
frozen system (i.e., $m^{*}=\infty$). We see that the ordering behavior of the
Ising model is essentially independent of mass \cite{Ising} and that the
results for the dynamically disordered systems lie very close to  those for
the randomly frozen case. Reduced constant volume heat capacities per
particle, $C_{V}/Nk$, obtained by numerically differentiating the average
dipolar energy with respect to the rotational temperature are also shown in
Fig. 7 (see inset)~\cite{comment_on_Cv}. The randomly frozen and $m^{*}=5$
results are very similar and indicate a phase transition at $T^{*} \approx
25$.

The dependence of the ferroelectric transition temperature, $T^{*}_{F}$
(rotational), on particle mass $m^{*}$ for the XY and XYZ models is shown in
Fig. 8. The transition temperatures were estimated from heat capacities (see
Fig. 8, inset) obtained from numerical differentiation of the dipolar energy
per particle. Results for the Ising system are not plotted because the
transition temperature is essentially independent of the mass \cite{Ising}. As
the mass increases the transition temperature drops for both the XY and XYZ
models. As noted earlier, for large masses and low rotational temperatures
convergence becomes prohibitively slow, but it seems reasonable to assume that
the graph would simply continue with the transition temperature approaching
zero in the infinite mass limit. This is clearly consistent with the fact that
we did not find a ferroelectric phase for randomly frozen systems at finite
temperatures.

\section{Mean Field Theory: Isolating the Effect of Long-Range
Interactions}

In order to better understand the simulation results discussed above, it is
useful to consider a simple mean field theory for a system where only
long-range reaction field interactions are included. The properties of such a
system are completely independent of spatial structure and it is instructive
to compare its behavior with the simulation results for the various models.

We consider $N$ particles in a spherical cavity of radius, $r_{c}$, surrounded
by a continuum of dielectric constant $\epsilon^{'}$.
The reaction field, ${\bf R}$, within the
cavity arises from the polarization of the surroundings by the dipoles in the
sphere and is given by \cite{allen,bottcher}
\begin{mathletters}
\begin{equation}
{\bf R}  = f(\epsilon'){\bf M}/r_{c}^{3}\ ,
\end{equation}
where
\begin{equation}
{\bf M} = \sum_{i=1}^{N} \bbox{\mu}_{i}\ ,
\end{equation}
\end{mathletters}
is the total dipole moment of the sphere and $f(\epsilon')=2(\epsilon'-1)/
(2\epsilon' +1)$. If, as in the simulations, $\epsilon'=\infty$, then
$f(\epsilon')=1$ and we shall write all further expressions for this specific
case. It is also obvious from the definition of $f(\epsilon')$, that as
$\epsilon'$ increases $f(\epsilon')$ rapidly approaches 1 and becomes
effectively independent of the exact value of $\epsilon'$. This is the
justification mentioned above for using $\epsilon'=\infty$ in simulations of
dipolar systems which large dielectric constants.

The total instantaneous energy of the system, $U$, can be written in the
form
\begin{equation}
U=-\frac{1}{2}\sum_{i=1}^{N}\bbox{\mu}_{i} \cdot {\bf R}\ ,
\end{equation}
where factor of $1/2$ arises when one calculates the energy of a point dipole
in a reaction field \cite{allen,bottcher,frohlich}. It takes account of the
work required to polarize the surrounding continuum. Clearly, the low energy
state is when the dipoles are aligned with the reaction field. Using Eqs (7),
$U$ can be expressed in the form
\begin{equation}
U=-\frac{2\pi}{3}\rho\mu^{2}\frac{1}{N}\sum_{i=1}^{N}\sum_{j=1}^{N}
\hat{\bbox{\mu}}_{i} \cdot \hat{\bbox{\mu}}_{j}\ ,
\end{equation}
where we have also used $N=4\pi r_{c}^{3}\rho/3$ and $\rho=N/V$.

In the canonical ensemble the average polarization can be obtained by
evaluating
\begin{equation}
P = \frac{\int P_{\rm ins}e^{-\beta U} d{\bf \Omega}^{N}}
    {\int e^{-\beta U} d{\bf \Omega}^{N}}\ ,
\end{equation}
where
$P_{\rm ins} =(1/N) \sum_{i=1}^{N} \hat {\bbox{\mu}}_{i} \cdot \hat {\bf d}$ is
the instantaneous polarization and $d{\bf \Omega}^{N}$ represents integration
over the angular coordinates of the N particles (here $\hat{{\bf d}}$ defines
the $z$ axis of the coordinate system). If we make the mean field
approximation \cite{mfref}
\begin{equation}
(\hat{{\bbox{\mu}}}_{i} - P\hat{{\bf d}})\cdot
(\hat{{\bbox{\mu}}}_{j} - P\hat{{\bf d}}) =0\ ,
\end{equation}
then Eq. (9) simplifies to
\begin{equation}
U \approx -\frac{4\pi}{3}\rho\mu^{2}P\sum_{i=1}^{N} \hat{\bbox{\mu}}_{i}
\cdot \hat{{\bf d}} + \frac{2\pi}{3}\mu^{2}NP^{2}\ ,
\end{equation}
and Eq. (10) can be easily solved for $P$.
Actually, here we dealing with an effective infinite range system where each
dipole is coupled with every other dipole with the coupling constant
$2\pi\rho\mu^{2}/3N$ [see Eq. (9)]. In this infinite range case, the mean
field solution is equivalent to the Hubbard-Stratonovich solution and is
essentially exact \cite{HS}.
For the XYZ, XY and Ising models, respectively, one obtains
\begin{mathletters}
\begin{equation}
P = \coth(x) - 1/x\ ,
\end{equation}
\begin{equation}
 P = \frac{ {\rm I}_1(x)}{ {\rm I}_0(x)} \ , 
\end{equation}
\begin{equation}
 P = {\rm tanh}(x) \ ,
\end{equation}
\end{mathletters}
where $x=4 \pi \rho \beta \mu^2 P/3$ and I$_n$(x) is the modified Bessel
function of order $n$. The average energy per particle is given by
\begin{equation}
\langle U \rangle/N = -2\pi \rho \mu^{2} P^{2}/3\ ,
\end{equation}
and the constant volume heat capacity can be obtained by taking the
differentiation with respect to temperature.

The mean field theory predicts ferroelectric
transitions for all three models. By expanding Eqs (13) about $x=0$ one
finds that the mean field ferroelectric transition temperatures are given
by
\begin{equation}
T^{*}_{F} = 4\pi\rho^{*}{\mu^{*}}^{2}/3n\ ,
\end{equation}
where $n$ is the number of dipole components. For the Ising and XYZ models Eq.
(15) agrees with the ``conventional mean field'' results given by Zhang and
Widom \cite{zw95}. For the present parameters ($\mu^{*}=4$, $\rho^{*}=0.8$) the
transition temperatures obtained are 17.9, 26.8 and 53.6 for the XYZ, XY and
Ising models, respectively. We emphasize that the driving force for the
transition in this simple theory is just the reaction field due to the
self-consistent polarization of the surrounding continuum.

It is clear from the above discussion that the reaction field interactions
favor ferroelectric order for all three models. However, as described in
Sections IV and V, computer simulations of randomly frozen or dynamically
decoupled systems indicate the existence of ferroelectric order only for the
Ising model. Ferroelectric phases were not observed for the XY and XYZ models
and the MD evidence strongly suggests that these systems are unpolarized at
all finite temperatures. This behavior can be explained if we consider that
the reaction field, ${\bf R}$, is only one contribution to the total local
field, ${\bf E}_{{\rm local}}$, experienced by a particle. The local field can
be written as
\begin{equation}
{\bf E}_{{\rm local}} = {\bf R} + {\bf E}\ ,
\end{equation}
where the remaining contribution, ${\bf E}$, is dependent on positional
correlations and may or may not favor ferroelectric order. If ${\bf R}$
dominates, ferroelectric phases are to be expected. However, if the local
field is largely determined by ${\bf E}$, then the existence, or non-existence
of ferroelectric phases will depend on the details of the spatial
correlations (eg., as in Bravais lattices).

>From this point of view, it is useful to divide the average energy obtained
in the simulations into reaction field (RF) and structurally dependent
(SD) parts such that
\begin{equation}
\langle U \rangle = \langle U_{\rm RF} \rangle + \langle U_{\rm SD} \rangle\ .
\end{equation}
Of course, $\langle U_{\rm RF} \rangle$ depends only on particle orientation
with respect to ${\bf R}$, whereas $\langle U_{\rm SD} \rangle$ will be
sensitive to details of the local spatial structure experienced by a particle.
The total energy and both contributions obtained in dynamically decoupled
simulations of all three models are shown as functions of $T^{*}$ (rotational)
in Fig. 9. Results for different values of the reduced mass are included. For
all three models we see that $\langle U_{\rm RF} \rangle$ dominates in the
ferroelectric state and that $\langle U_{\rm SD} \rangle$ dominates in the
paraelectric phase. In fact, in all cases, as the system becomes ferroelectric
$\langle U_{SD} \rangle$ increases (i.e., becomes less negative) indicating
that the structurally dependent interactions do not favor ferroelectric order.

For the XYZ (Fig. 9a) and XY (Fig. 9b) models the different contributions to
the energy are strongly mass dependent. As the mass increases, and the
particles move more and more slowly, $\langle U_{\rm SD} \rangle$ dominates at
lower and lower rotational temperatures. Thus, if we begin at a fixed
rotational temperature in the ferroelectric phase and increase the mass, the
dipoles eventually respond to the details of the random local structure,
$\langle U_{\rm SD} \rangle$ dominates, and the ferroelectric phase disorders.
The Ising model does not exhibit this behavior. In the Ising case both
contributions to the energy are largely independent of mass (Fig. 9c) and a
stable, mass-independent, ferroelectric phase is observed. The difference in
behavior of the Ising model must arise from the fact that, with only two
dipolar directions available, its response to local structure is much more
limited than that possible for the XY and XYZ systems.

\section{Summary and Conclusions}

In this paper, we have used computer simulation methods to explore
the phase behavior of spatially disordered dipolar systems. Both randomly
frozen and dynamically disordered models were considered. It was found that
the behavior observed depended upon the number of components included in
the dipoles. Ferroelectric phases were not observed for the randomly
frozen XY and XYZ models. Rather, these systems appear to freeze
orientationally into dipolar glasses as the temperature is lowered. The
behavior of the randomly frozen Ising model was significantly different.
This model does have a stable ferroelectric phase at sufficiently low
temperatures.

We also considered dynamically disordered systems simulated such that the
translational motion of the particles was independent of the dipolar forces.
The rate of translational motion was adjusted by varying the particle mass.
Interacting dipoles embedded in this moving substrate can respond to the
short-range spatial structure only if the motion is sufficiently slow. For
small masses (rapid translational motion), ferroelectric phases were observed
for all three models. However, as the mass is increased and the dipoles become
``aware'' of the short-range random structure, the ferroelectric order is
destroyed in the XY and XYZ models. In the infinite mass limit both the XY and
XYZ models appear to be paraelectric at all temperatures. This is consistent
with our observations for the randomly frozen systems. In contrast, the Ising
model exhibits a paraelectric-to-ferroelectric transition at a temperature
which is essentially independent of mass and agrees with the result for the
randomly frozen case.

In order to understand our observations, it is useful to divide the total
local field experienced by a dipole into two parts. These are, a reaction
field contribution which has no dependence on the local spatial structure,
and a structure-dependent part which includes everything else. Clearly,
the average total energy can be divided into corresponding reaction field and
structure-dependent contributions. Both contributions to the total energy
were evaluated in our simulations. In all systems where ferroelectric order
exists, it was shown to be stablized by the reaction field interactions.
In spatially random systems, the structure-dependent contribution never
favors ferroelectric order. In the XY and XYZ models, the structure-dependent
part dominates and paraelectric dipolar glasses rather than ferroelectric
phases are observed. For the Ising model, the reaction field contribution
dominates and ferroelectric order is observed. In all likelihood, the Ising
model differs from the XY and XYZ systems simply because with a one-component
dipole the opportunities for strong local interactions are severely limited.

\acknowledgments

We thank Z. R\'acz, M. Widom and  H. Zhang for useful discussions. Also, we
are much indepted to M. Widom and Shuboho Banerjee for pointing out an error
in an earlier manuscript version of this article. The financial support of the
National Science and Engineering Research Council of Canada is gratefully
acknowledged.

\begin{figure}
\caption{
A sketch of a possible phase diagram for spatially disordered dipolar
systems. The various terms, symbols and lines are referred to in the
text.}
\end{figure}

\begin{figure}
\caption{
The radial distribution function, $g(r)$, for soft spheres at $\rho = 0.8$
and $T^{*} = 10.5$.}
\end{figure}

\begin{figure}
\caption{
The polarization $P$ at $T^{*}_{{\rm min}}$ vs $100/N$ for the randomly
frozen Ising (circles), XY (squares) and XYZ (triangles) models. Results are
included for 108 (XYZ only), 256, 500 and 864 particles.}
\end{figure}

\begin{figure}
\caption{
The root-mean-square dipole length, $\langle S_{\rm rms} \rangle$, for the frozen XYZ
model with $N=256$.}
\end{figure}

\begin{figure}
\caption{
$P$ vs $T^{*}$ (rotational) for dynamically random XYZ
systems. The squares, triangles and circles are for $m^{*}=1$, 5 and 10,
respectively. The error bars represent one estimated standard deviation.
$g_{\rm Binder}$ vs $T^{*}$ (rotational) is shown in the inset for $N=64$
(squares), 108 (triangles) and 256 (circles) particles.}
\end{figure}

\begin{figure}
\caption{
$P$ vs $T^{*}$ (rotational) for dynamically random XY
systems. The squares, triangles and circles are for $m^{*}=1$, 5 and 10,
respectively. The error bars represent one estimated standard deviation.}
\end{figure}

\begin{figure}
\caption{
$P$ vs $T^{*}$ (rotational) for the Ising model. Results are shown for
dynamically random systems with $m^{*}=1$ (squares) and $m^{*}=5$ (triangles)
and for the randomly frozen case (solid circles). The reduced heat capacities per
particle, $C_{V}/Nk$, are plotted vs $T^{*}$ (rotational) in the inset.}
\end{figure}

\begin{figure}
\caption{
The mass dependence of the disordered-to-ferroelectric transition temperature
$T^{*}_{F}$ (rotational). The squares and triangles are results for the XY
and XYZ models, respectively. The values of $T^{*}_{F}$ were obtained from
plots of the heat capacity, $C_{V}/Nk$, vs $T^{*}$ (rotational) and a
typical example is shown in the inset. The error bars represent estimated
uncertainties in the peak position.}
\end{figure}

\begin{figure}
\caption{
The contributions to the average energy for the XYZ (a), XY (b) and Ising (c)
models. $T^{*}$ is the rotational temperature. The solid, crossed and open
symbols denote $\langle U \rangle$, $\langle U_{\rm SD} \rangle$ and $\langle
U_{\rm RF} \rangle$, respectively. The squares, triangles and circles indicate
$m^{*}=1$, 5 and 10, respectively.}
\end{figure}


\begin{references}
\bibitem{restric_to_FE}For compactness, we shall refer here only to
ferroelectric order, but the discussion and results apply to ferromagnetic
order of dipolar origin as well.
\bibitem{Ising} We note that other decoupled simulation schemes could be
used for the Ising model. For example, if 100 MD timesteps are performed
between MC sweeps mass dependence is observed with the transition moving
to higher temperatures for lighter masses. However, in the large mass
limit one again recovers the results shown in Fig. 7.
\bibitem{comment_on_Cv} Here, as in Levy flight systems~\cite{levy}, we are
dealing with non-equilibrium conditions. Therefore, the heat capacity obtained
from the temperature derivative of the energy need not and does not agree with
that estimated from energy fluctuations.
\end{references}
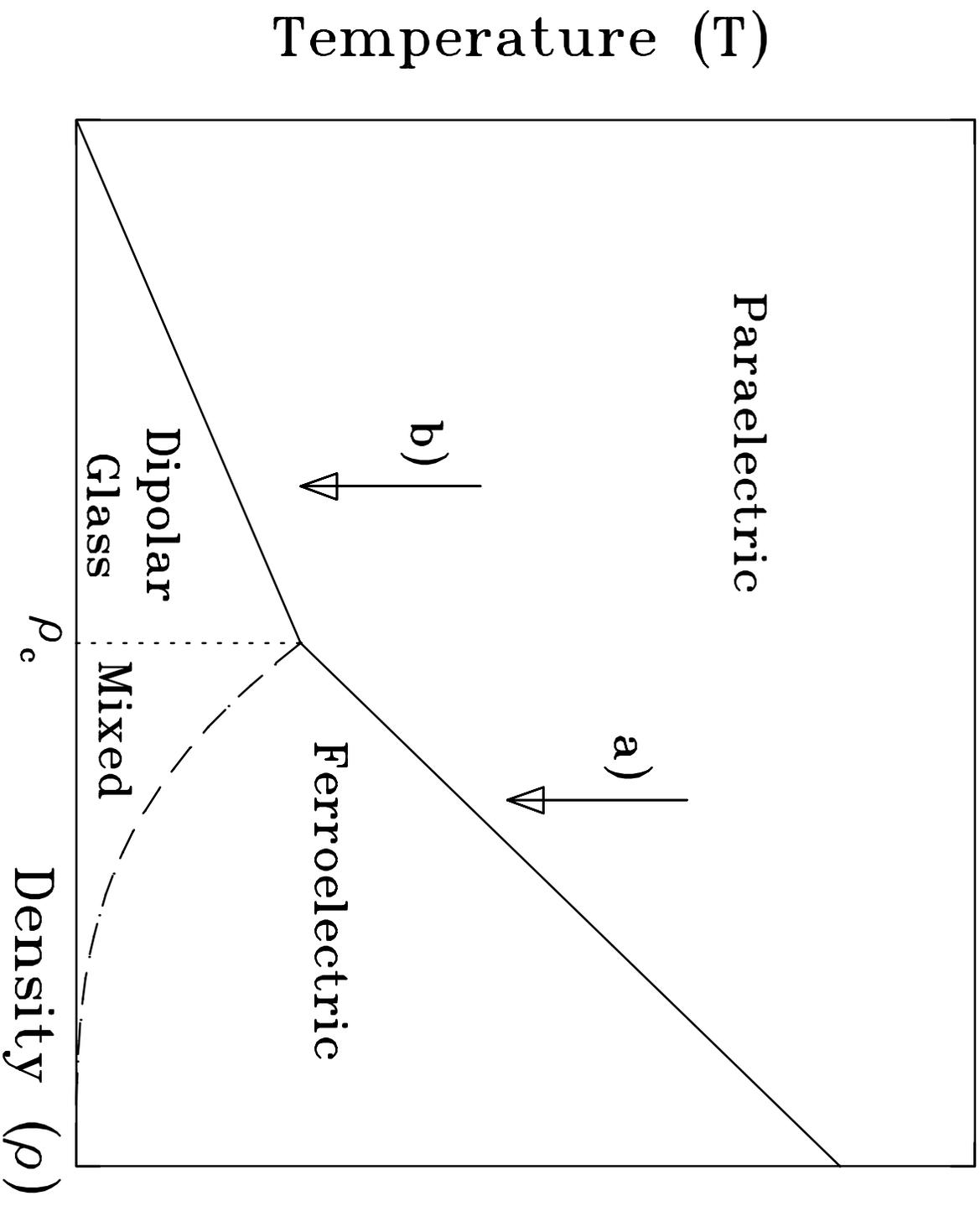
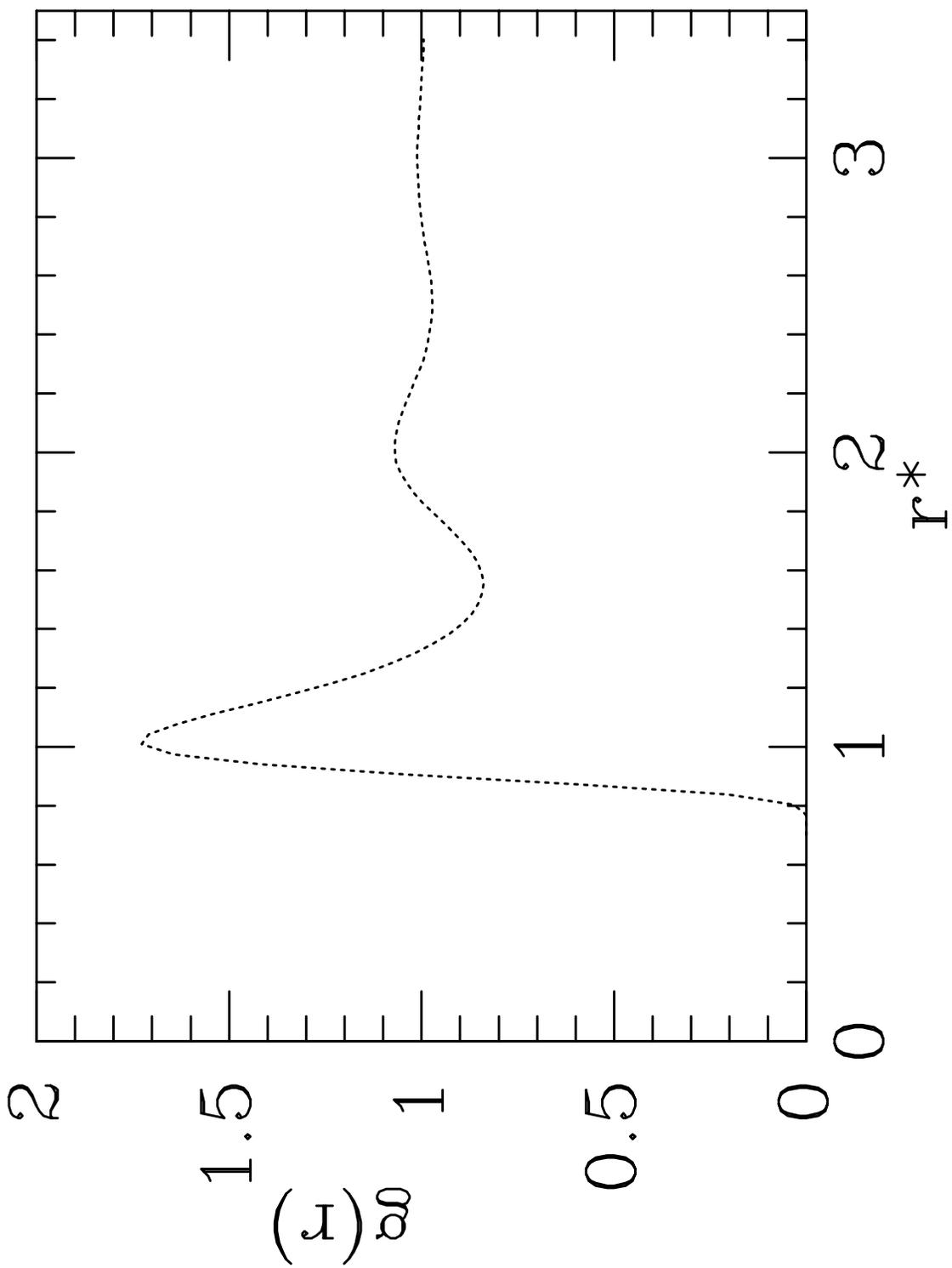
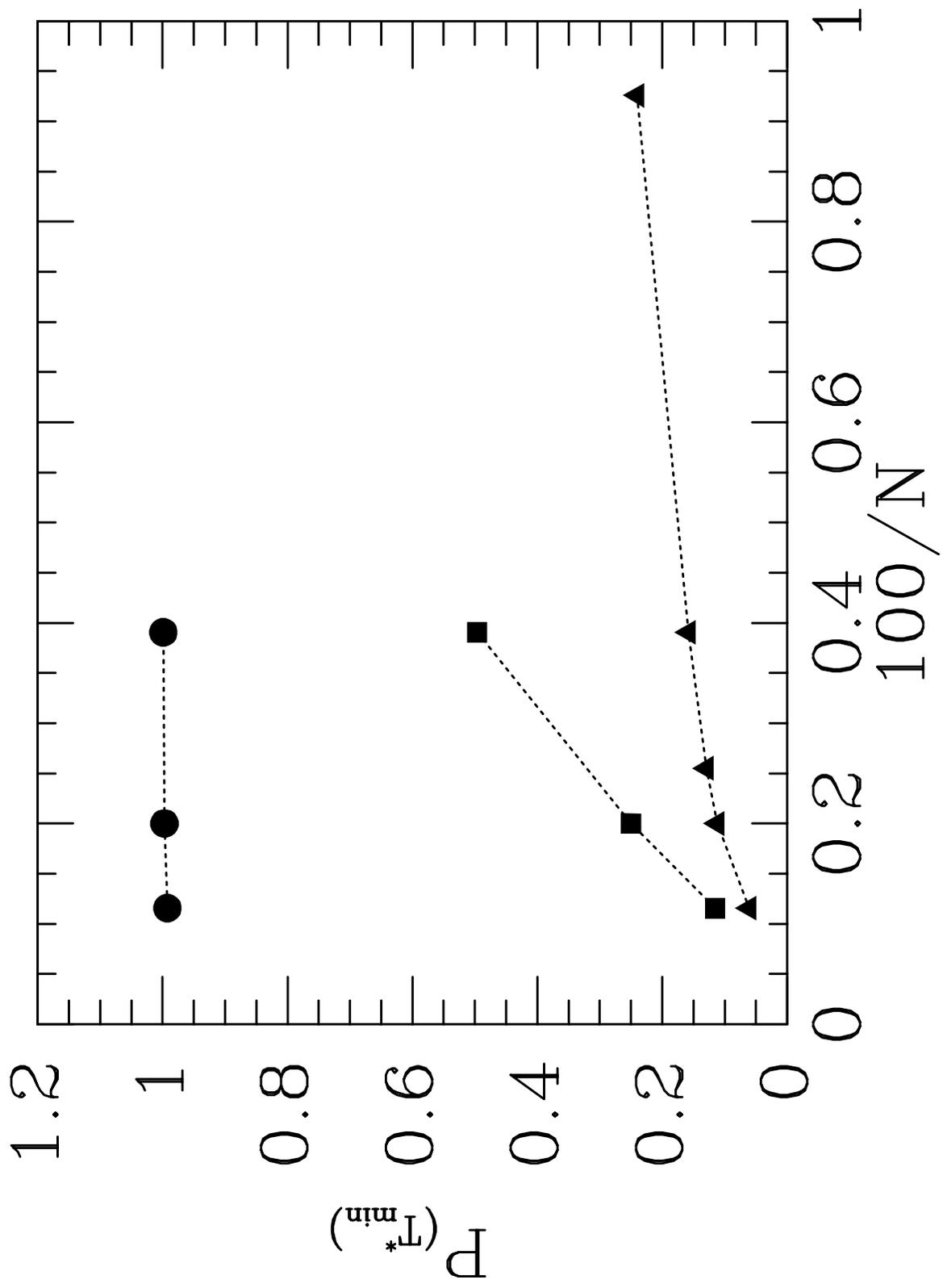
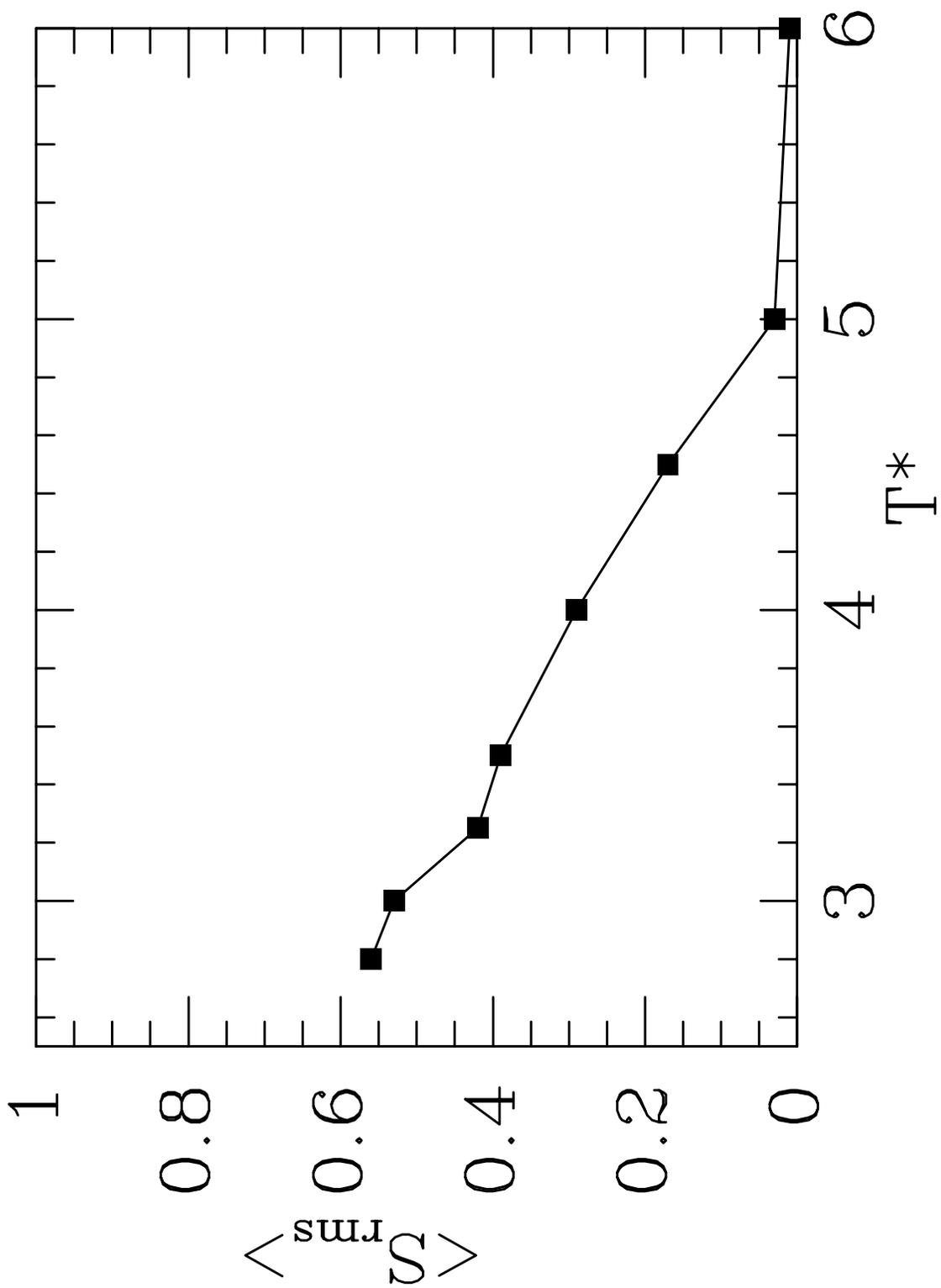
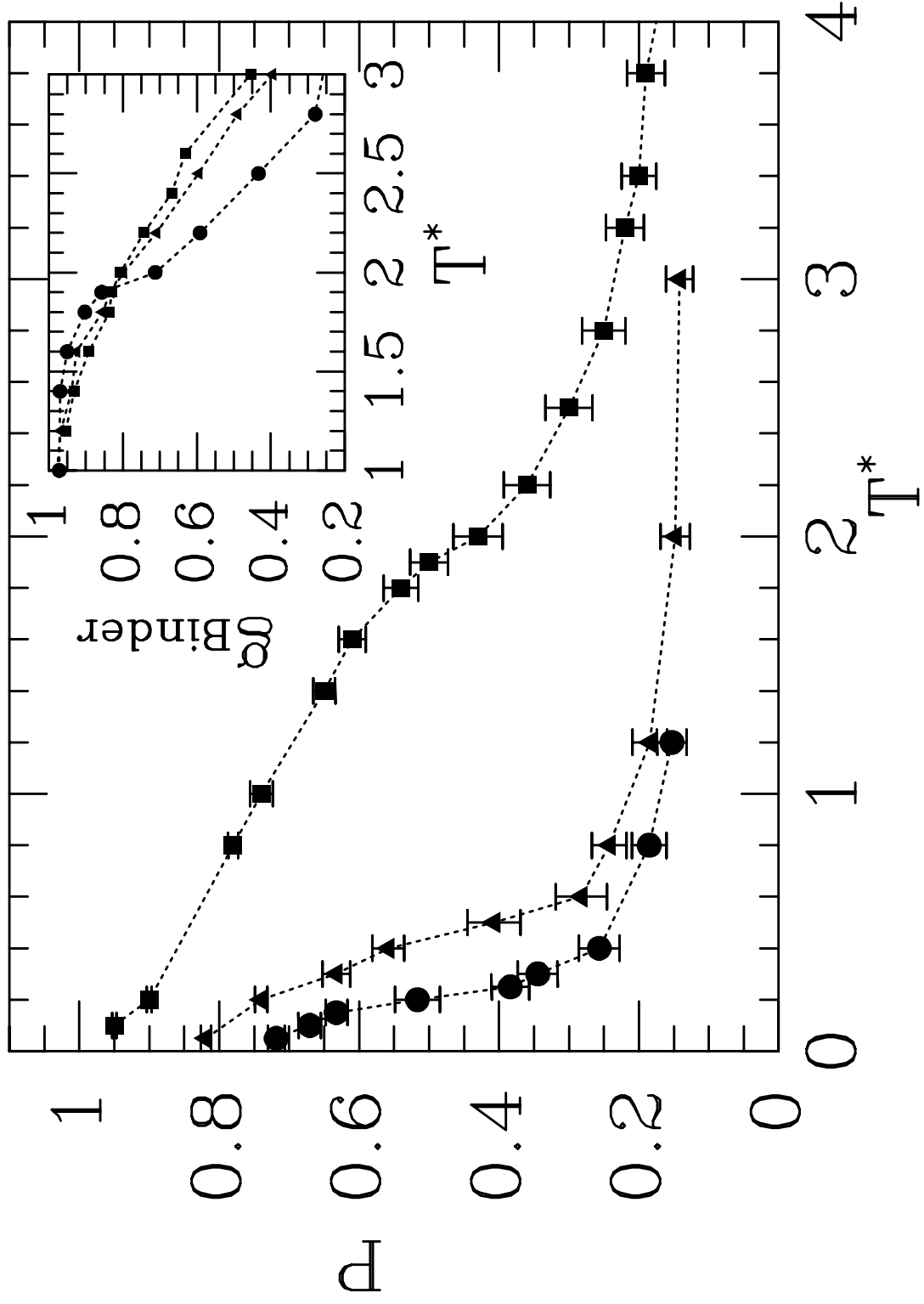
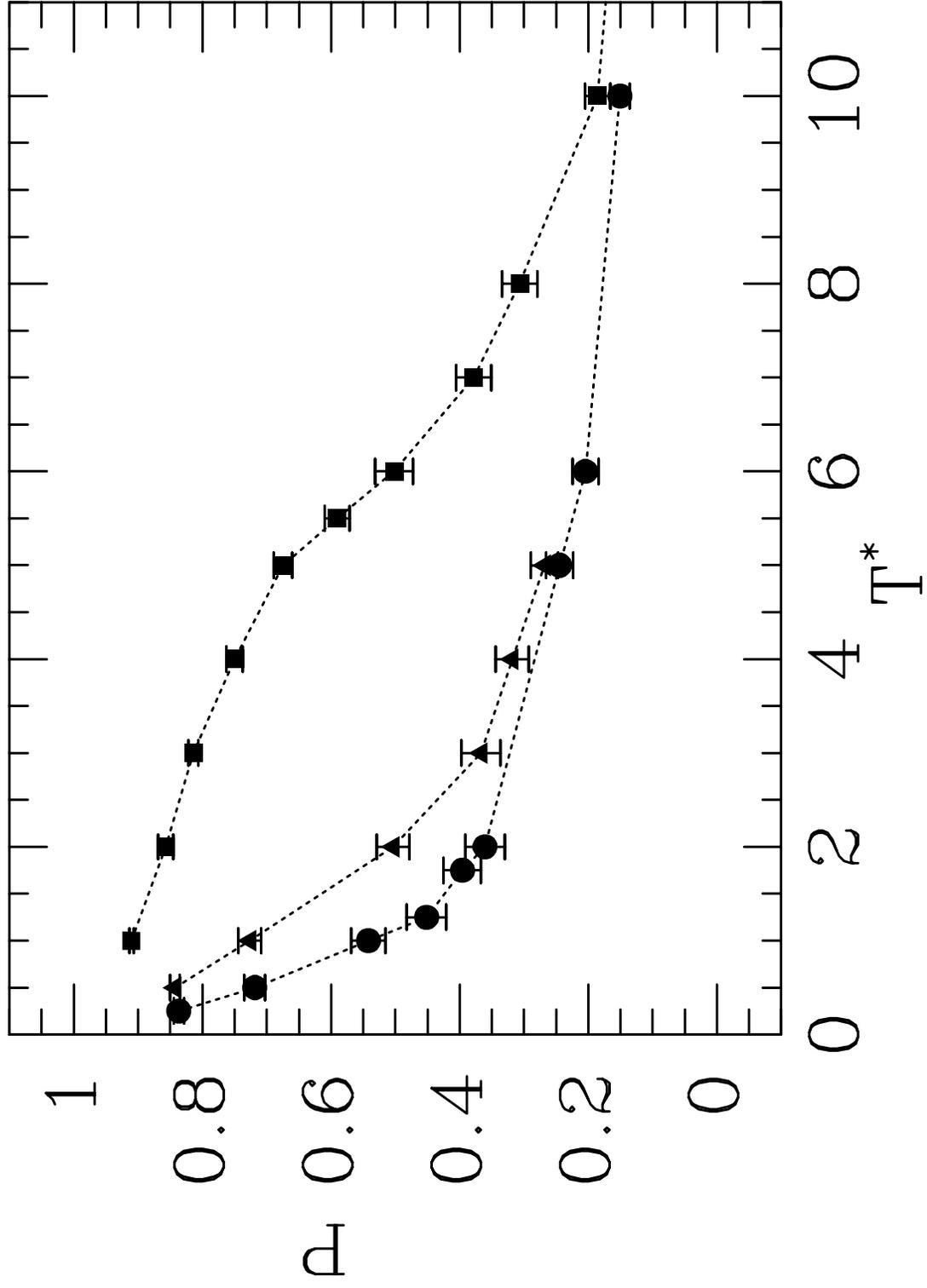
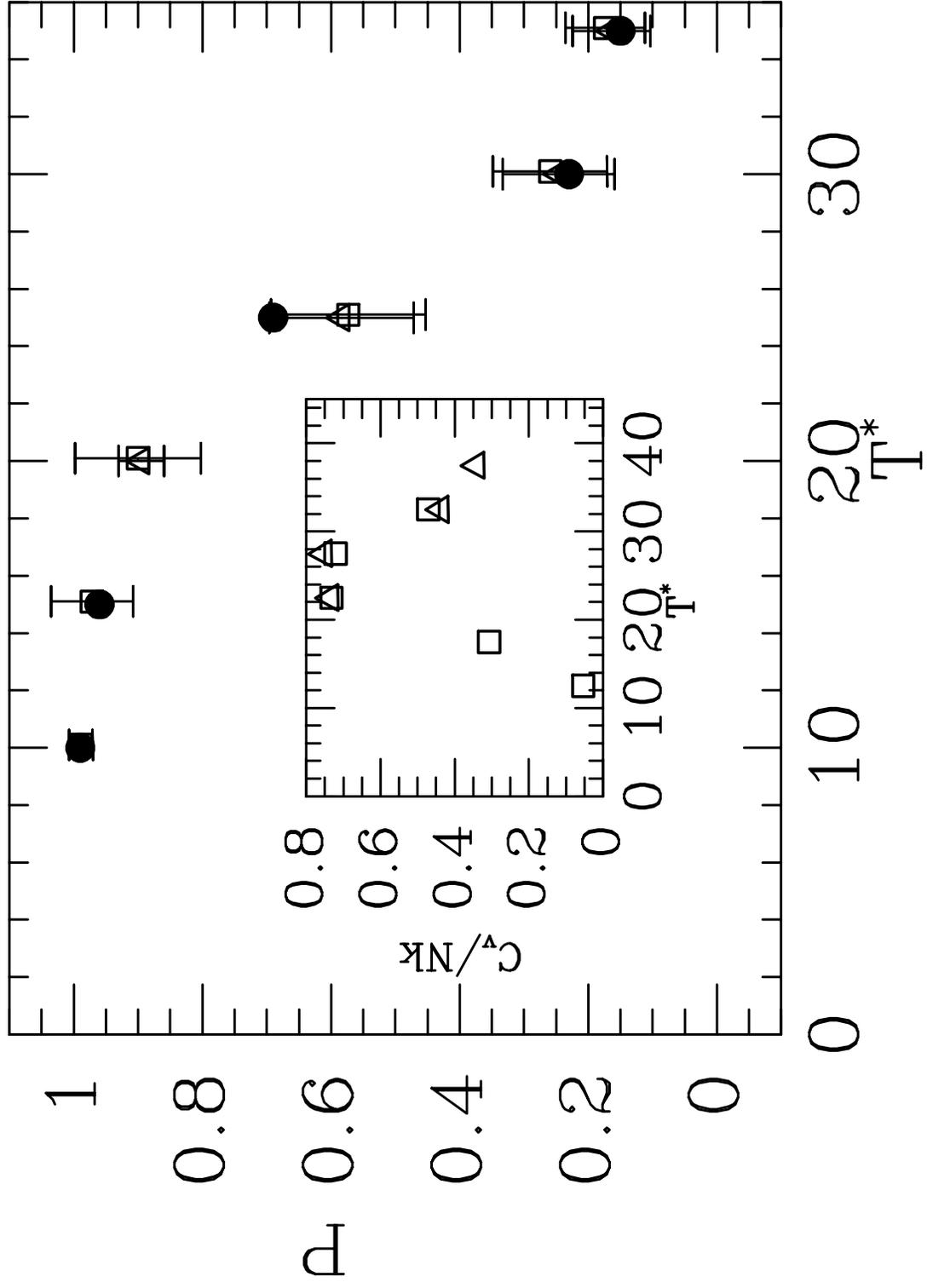
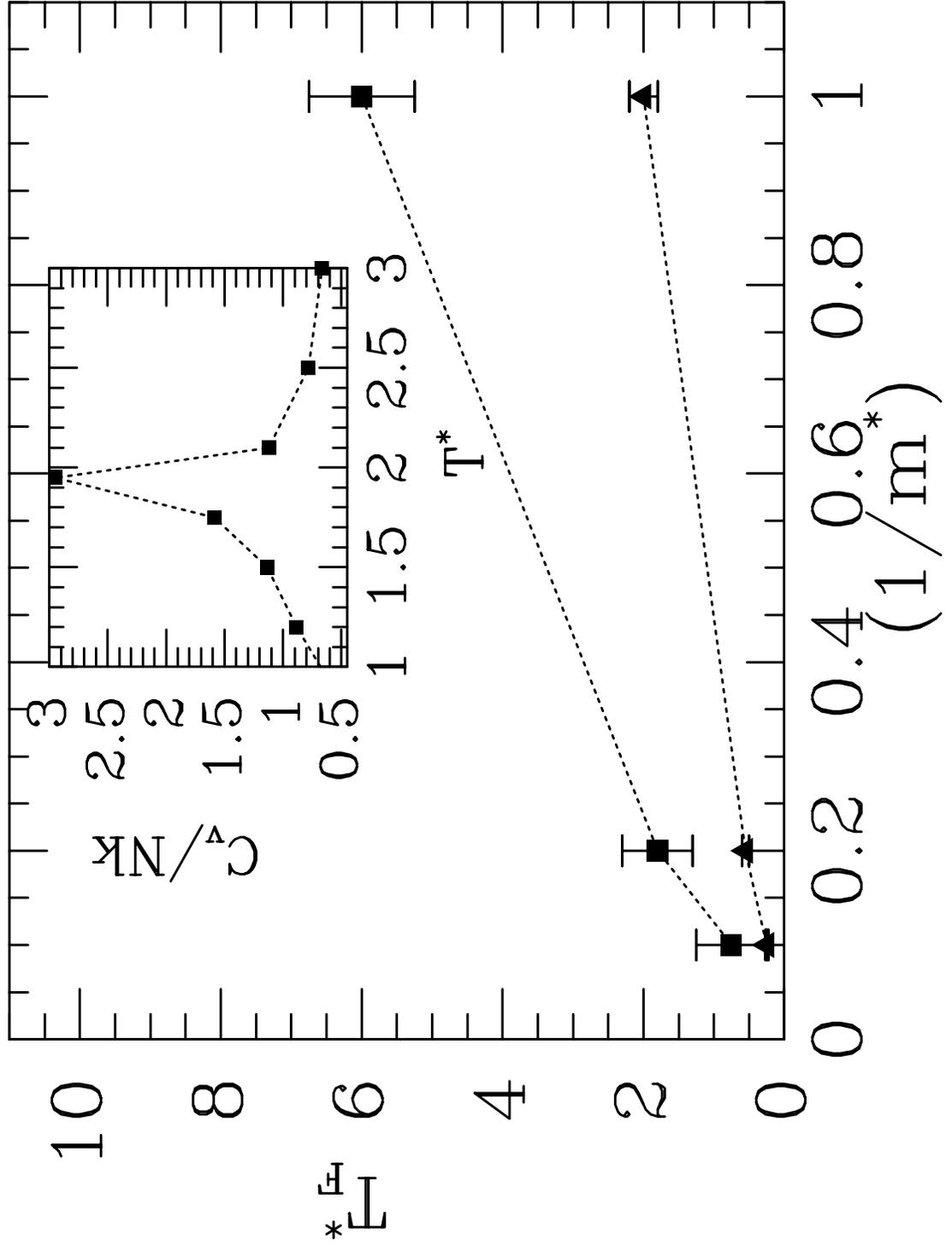
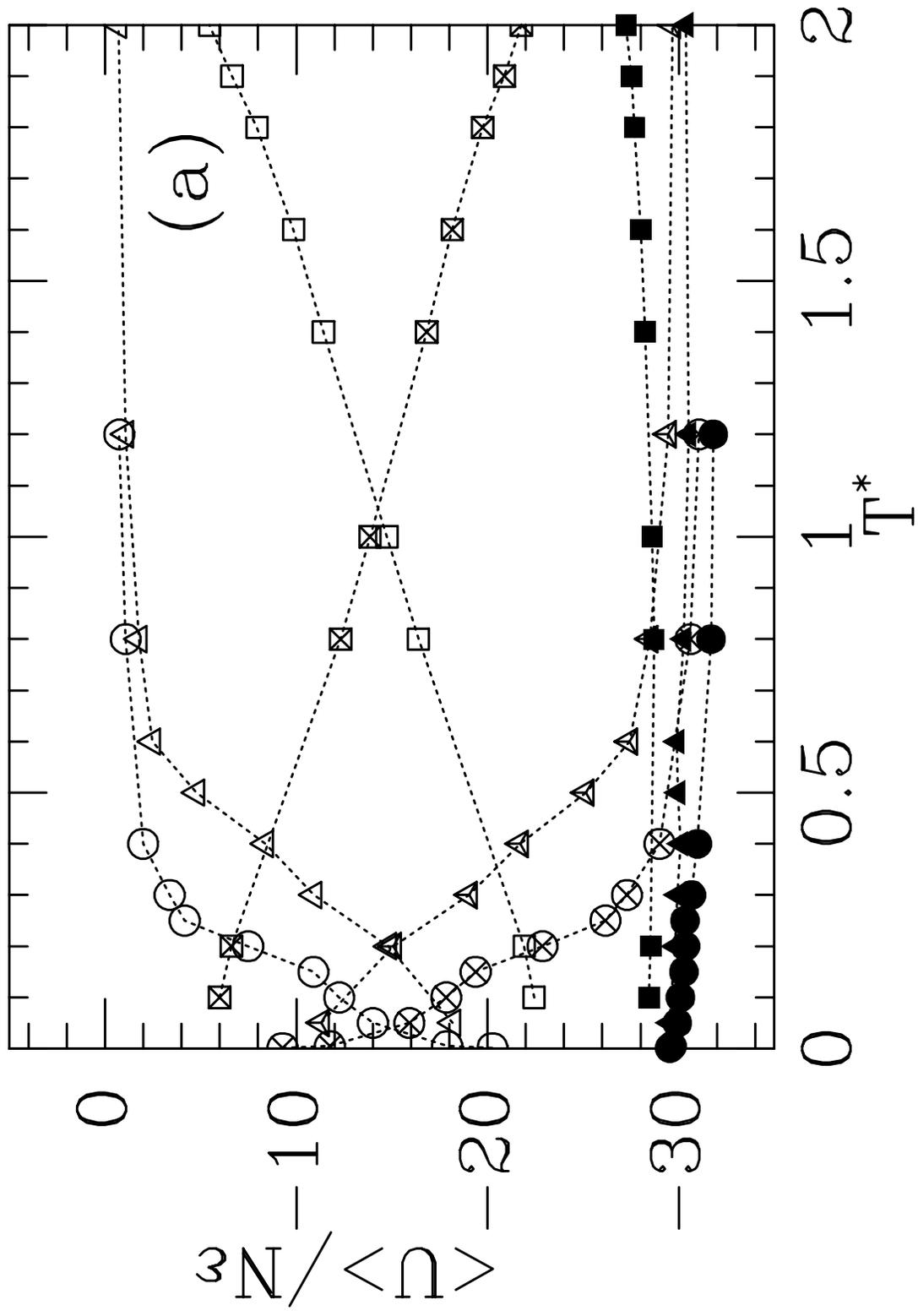
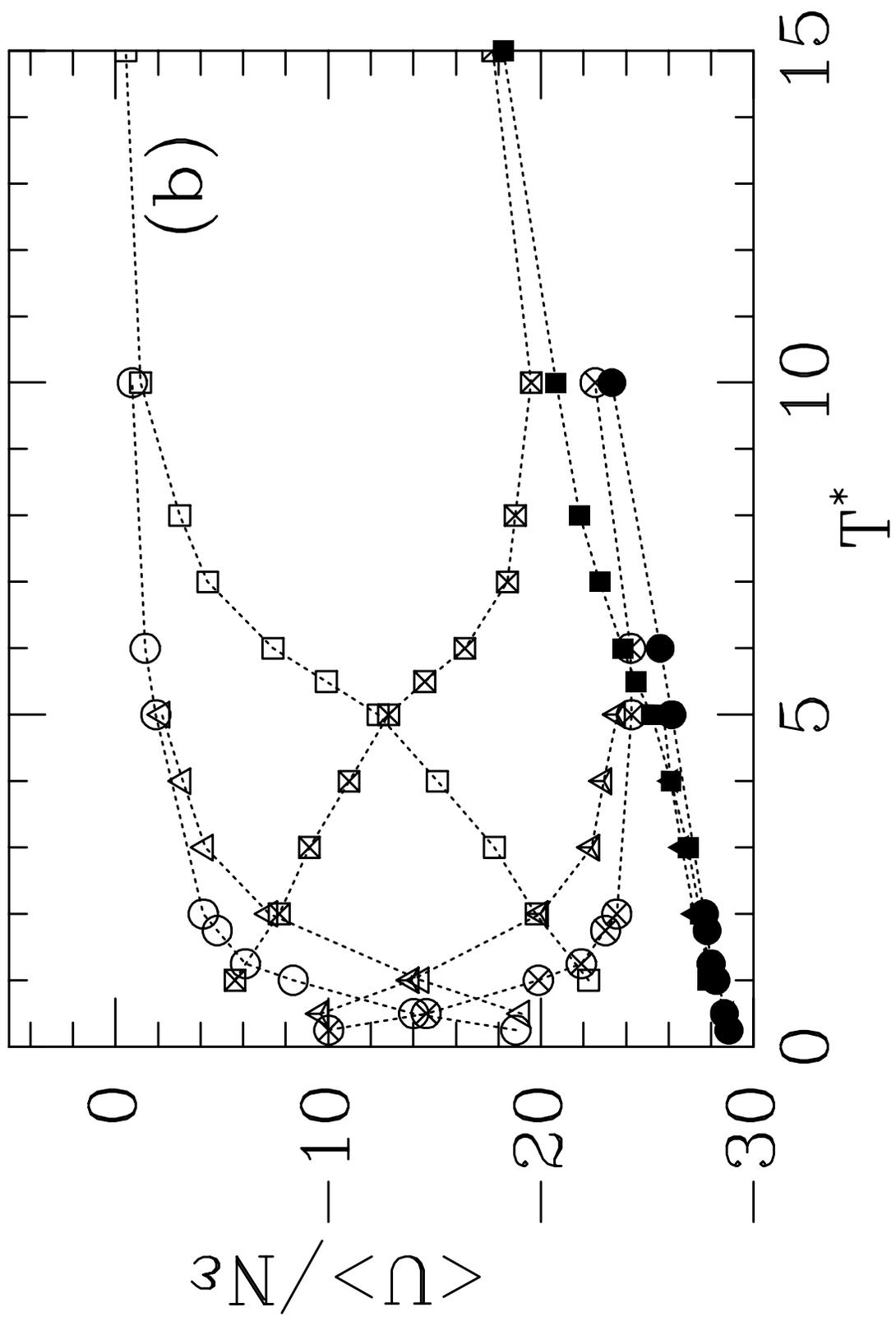
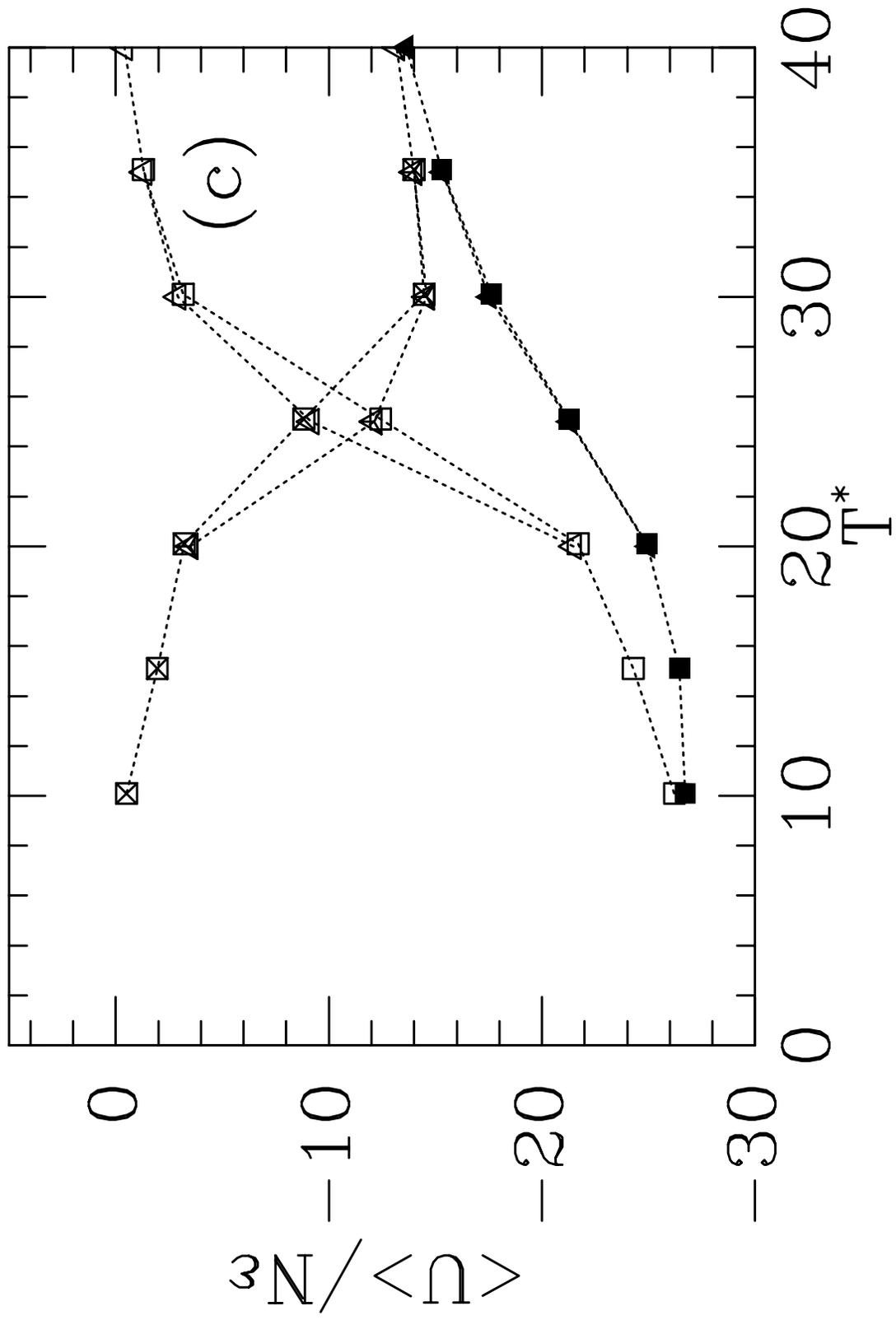
\end{document}